\documentstyle[11pt,newpasp,twoside,epsf]{article}
\def\edcomment#1{\iffalse\marginpar{\raggedright\sl#1\/}\else\relax\fi}
\reversemarginpar

\begin{document}
\title{Three-Dimensional Studies of the Warm Ionized Medium in the
Milky Way using WHAM}
 \author{R. J. Reynolds, L. M. Haffner, and G. J. Madsen}
\affil{University of Wisconsin-Madison, Astronomy Department, 475 N. 
Charter St., Madison, WI 53706}

\begin{abstract}
 The Wisconsin H-Alpha Mapper (WHAM) is a high throughput Fabry-Perot
facility developed specifically to detect and explore the warm, ionized
component of the interstellar medium at high spectral resolution.  It
began operating at Kitt Peak, Arizona in 1997 and has recently completed
the WHAM Northern Sky Survey (WHAM-NSS), providing the first global view
of the distribution and kinematics of the warm, diffuse H~II in the Milky
Way.  This H$\alpha$ survey reveals a complex spatial and kinematic
structure in the warm ionized medium and provides a foundation for studies
of the temperature and ionization state of the gas, the spectrum and
strength of the ionizing radiation, and its relationship to other
components of the interstellar medium and sources of ionization and
heating within the Galactic disk and halo.
\end{abstract}

\section{Introduction}

Warm ionized gas is a principal component of the interstellar medium in
our Galaxy and others.  Its large scale height, mass surface density, and
power requirement have significantly modified our understanding of the
composition and structure of the interstellar medium and the distribution
and flux of ionizing radiation within the disk and halo (e.g., Kulkarni \&
Heiles 1987; McKee 1990, Reynolds 1991, Ferri\`{e}re 2001).  Although
originally detected in the 1960s with radio techniques, subsequent
developments in high-throughput Fabry-Perot spectroscopy have shown that
the primary source of information about the distribution, kinematics, and
other physical properties of this gas is obtained through the detection
and study of faint, diffuse interstellar emission lines at optical
wavelengths.  Presented below are some recent results from the Wisconsin
H$\alpha$ Mapper (WHAM), including velocity-interval maps from the
recently completed WHAM Northern Sky Survey (WHAM-NSS) of interstellar
H$\alpha$ as well as observations of much fainter ``diagnostic'' emission
lines that probe the ionization and excitation state of the gas.

\subsection{The Warm Ionized Medium in the Milky Way}

Diffuse ionized gas is a major, yet poorly understood component of the
interstellar medium, which consists of regions of warm (10$^{4}$ K),
low-density (10$^{-1}$ cm$^{-3}$), nearly fully ionized hydrogen that
occupy approximately 20\% of the volume within a 2 kpc thick layer about
the Galactic midplane (e.g., Haffner, Reynolds, \& Tufte 1999).  Near the
midplane, the space averaged density of H~II is less than 5\% that of the
H~I. However, because of its greater scale height, the total column
density of interstellar H~II along high Galactic latitude sight lines is
relatively large, 1/4 to 1/2 that of the H~I, and one kiloparsec above the
midplane, warm H~II may be the dominant state of the interstellar medium
(Ferri\`ere 2001; Reynolds 1991b).  The presence of this ionized medium
can have a significant effect upon the interstellar pressure near the
Galactic midplane (Cox 1989) and upon the dynamics of hot (10$^5$ --
10$^6$ K), ``coronal'' gas far above the midplane (e.g., Heiles 1990).  
Miller \& Cox (1993) have suggested that this gas is part of a wide
spread, warm intercloud medium, while in the McKee \& Ostriker (1977)
picture of the interstellar medium, this warm H~II is located in the outer
envelopes of H~I clouds, forming the boundary between the clouds and a
wide spread, hot (``coronal'') phase.

It is generally believed that the O stars, confined primarily to widely
separated stellar associations near the Galactic midplane, are somehow
able to account for this widespread gas, not only in the disk but also
within the halo, 1-2 kpc above the midplane.  However, the nature of such
a disk-halo connection is not clear.  For example, the need to have a
large fraction of the Lyman continuum photons from O stars travel hundreds
of parsecs through the disk seems to conflict with the traditional picture
of H~I permeating much of the interstellar volume near the Galactic plane.  
It has been suggested that ``superbubbles'' of hot gas, especially
superbubbles that blow out of the disk (``galactic chimneys''), may sweep
large regions of the disk clear of H~I, allowing ionizing photons from the
O stars within them to travel unimpeded across these cavities and into the
halo (e.g., Norman 1991).  Another possibility is that the Lyman continuum
radiation itself is able to carve out extensive regions of H~II through
low density portions of the H~I (e.g., Miller and Cox 1993), perhaps
creating photoionized pathways or ``warm H~II chimneys'' that extend far
above the midplane (Dove and Shull 1994; Dove, Shull, and Ferrara 2000).  
Although the existence of superbubbles has long been established (e.g.,
Heiles 1984), direct observational evidence that such cavities are
actually responsible for the transport of hot gas and ionizing radiation
up into the Galactic halo is very limited.  

Interestingly, even though the source of ionization is believed to be O
stars, the temperature and ionization conditions within the diffuse
ionized gas appear to differ significantly from conditions within
classical O star H~II regions.  For example, anomalously strong [S~II]
$\lambda$6716/H$\alpha$ and [N~II] $\lambda$6584/H$\alpha$, and weak
[O~III] $\lambda$5007/H$\alpha$ emission line ratios (compared to the
bright, classical H~II regions) indicate a low state of excitation with
few ions present that require ionization energies greater than 23 eV
(Haffner et al 1999; Rand 1997).  This is consistent with the low
ionization fraction of helium, at least for the helium near the midplane
(Reynolds \& Tufte 1995; Tufte 1997; Heiles et al 1996), which implies
that the spectrum of the diffuse interstellar radiation field that ionizes
the hydrogen is significantly softer than that from the average Galactic O
star population.  Rand (1997) has also reported lower helium ionization in
the H~II halo of the edge-on galaxy NGC 891. 

Furthermore, it has recently become apparent that O star photoionization
models fail to explain observed spatial variations in some of the line
intensity ratios.  For example, the models do not explain the very large
increases in [N~II]/H$\alpha$ and [S~II]/H$\alpha$ (accompanied by an
increase in [O~III]) with distance from the midplane or the observed
constancy of [S~II]/[N~II] (see discussions by Reynolds et al 1999,
Haffner et al 1999, and Collins \& Rand 2001). The data seem to require
the existence of a significant \emph{non-ionizing} source of heat that
overwhelms photoionization heating at low densities within the ionized
medium (Reynolds et al 1999; Collins \& Rand 2001; Otte et al 2001;  
Bland-Hawthorn, Freeman, \& Quinn 1997).  Proposed sources include the
dissipation of MHD turbulence, Coulomb interactions with cosmic rays,
magnetic reconnection, and photoelectric heating by a population of very
small grains (see Minter \& Spangler 1996; Reynolds et al 1999;
Weingartner \& Draine 2001).

\subsection{WHAM}

The Wisconsin H-Alpha Mapper (WHAM) is a remotely controlled observing
facility, funded by the National Science Foundation and dedicated to the
detection and study of faint optical emission lines from the diffuse
ionized gas in the disk and halo of the Milky Way (Tufte 1997; Reynolds et
al 1998b, Haffner 1998).  The WHAM facility consists of a 15 cm aperture
dual-etalon Fabry-Perot spectrometer (the largest used in astronomy)
coupled to a 0.6 m aperture siderostat, which provide a one-degree
diameter beam on the sky and produce a 12 km s$^{-1}$ resolution spectrum
across a 200 km s$^{-1}$ spectral window.  The spectral window can be
centered on any wavelength between 4800 \AA\ and 7300 \AA\ using a gas
(SF$_6$) pressure (optical index) control system and a filter wheel.  The
tandem etalons greatly extend the effective ``free spectral range'' of the
spectrometer, improve the shape of the response profile, and suppress the
multi-order Fabry-Perot ghosts, especially those arising from the
relatively bright atmospheric OH emission lines within the pass band of
the interference filter.  A high quantum efficiency (78\% at H$\alpha$),
low noise (3 e$^{-}$ rms) CCD camera serves as a multichannel detector,
recording the spectrum as a Fabry-Perot ``ring image'' without scanning
(e.g., Reynolds et al 1998b).

The construction and testing of this facility at the University of
Wisconsin was completed in September 1996, and WHAM began operating on
Kitt Peak in January 1997 (see Fig. 2). Since then, WHAM has been
successfully collecting data nearly every clear, dark-of-the-moon period.
It has completed as its first major mission a 37,565 spectra H$\alpha$
survey of the northern sky, which has provided the first map of the large
scale distribution and kinematics of diffuse interstellar H~II that is
comparable to earlier 21~cm surveys of H~I (\S 2, below). WHAM is now
beginning its second major mission, a comprehensive study of fainter,
diagnostic emission lines that trace the excitation and ionization
conditions within the gas as well as the strength and spectrum of the
ionizing radiation (\S 3 \& \S4, below).

\section{The WHAM Northern Sky H$\alpha$ Survey}

From 1997 January through 1998 September, WHAM obtained 37,565 spectra
with its 1 degree diameter beam covering the sky on a $0\fdg 98$/cosb
$\times~0\fdg 85$ grid ($\ell$,b) north of declination $-30\deg$. Figure 1
shows the beam covering pattern for a small portion of the sky survey.  
The observations were obtained in ``blocks'', with each block usually
consisting of 49 pointings made sequentially in a boustrophedonic raster
of seven beams in longitude and seven beams in latitude.  Each block took
approximately 30 minutes, and from one to twenty blocks were observed in a
night.  The absence or presence of block boundary features in the
completed survey map provide an excellent gauge of the systematic errors
associated with observations taken on different nights and different times
of the year. The radial velocity interval for the survey was limited to
$\pm 100$ km s$^{-1}$ with respect to the LSR. This range includes nearly
all of the interstellar emission at high latitudes except the H$\alpha$
associated with High Velocity H~I Clouds (HVCs), which by definition have
radial velocities $|$v$| >$ 80 km s$^{-1}$.

\begin{figure}

  \plotone{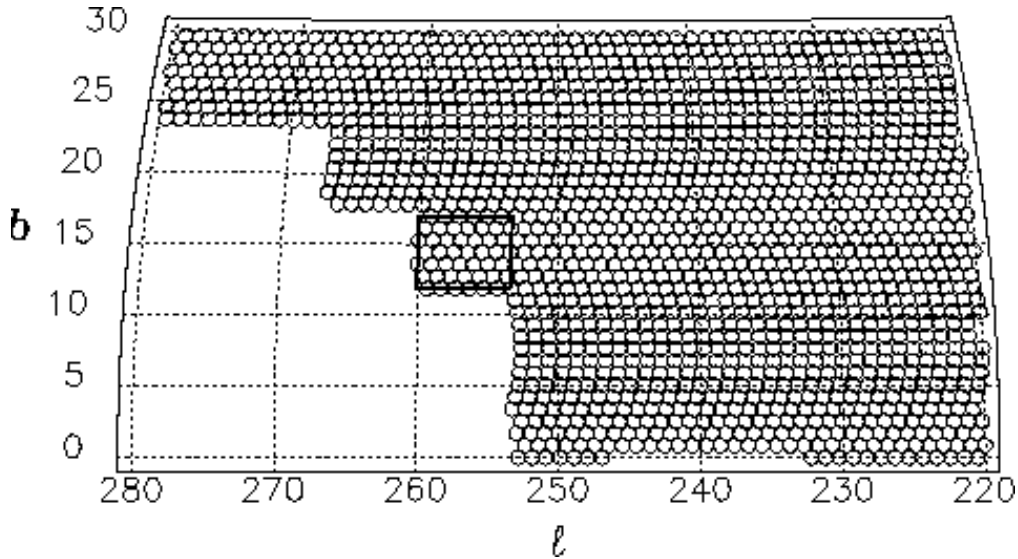}
  
  \caption{A portion of the sky near the southern declination limit 
    of the survey showing the pattern of WHAM's 1$\deg$ 
    diameter beams. Observations were obtained as a sequence of
    ``blocks'' (outlined), consisting typically of a grid of 49
    pointings within in a $7\deg \times 6\deg$ region. The integration  
    time per pointing was 30 s.}

\end{figure}

Figure 2 shows the resulting survey maps, including views of the total
H$\alpha$ intensity in addition to velocity interval maps.  Interstellar
H$\alpha$ emission extents over virtually the entire sky, with blobs and
filaments of enhanced H$\alpha$ superposed on a more diffuse background.  
The highest H$\alpha$ intensities are found near the Galactic equator,
with a general decrease toward the poles.  Some of the new features
revealed by this survey are discussed by Haffner (2001), Reynolds et al
(2001a), and Haffner, Reynolds, \& Tufte (1998).

\begin{figure}[htbp]
\plotone{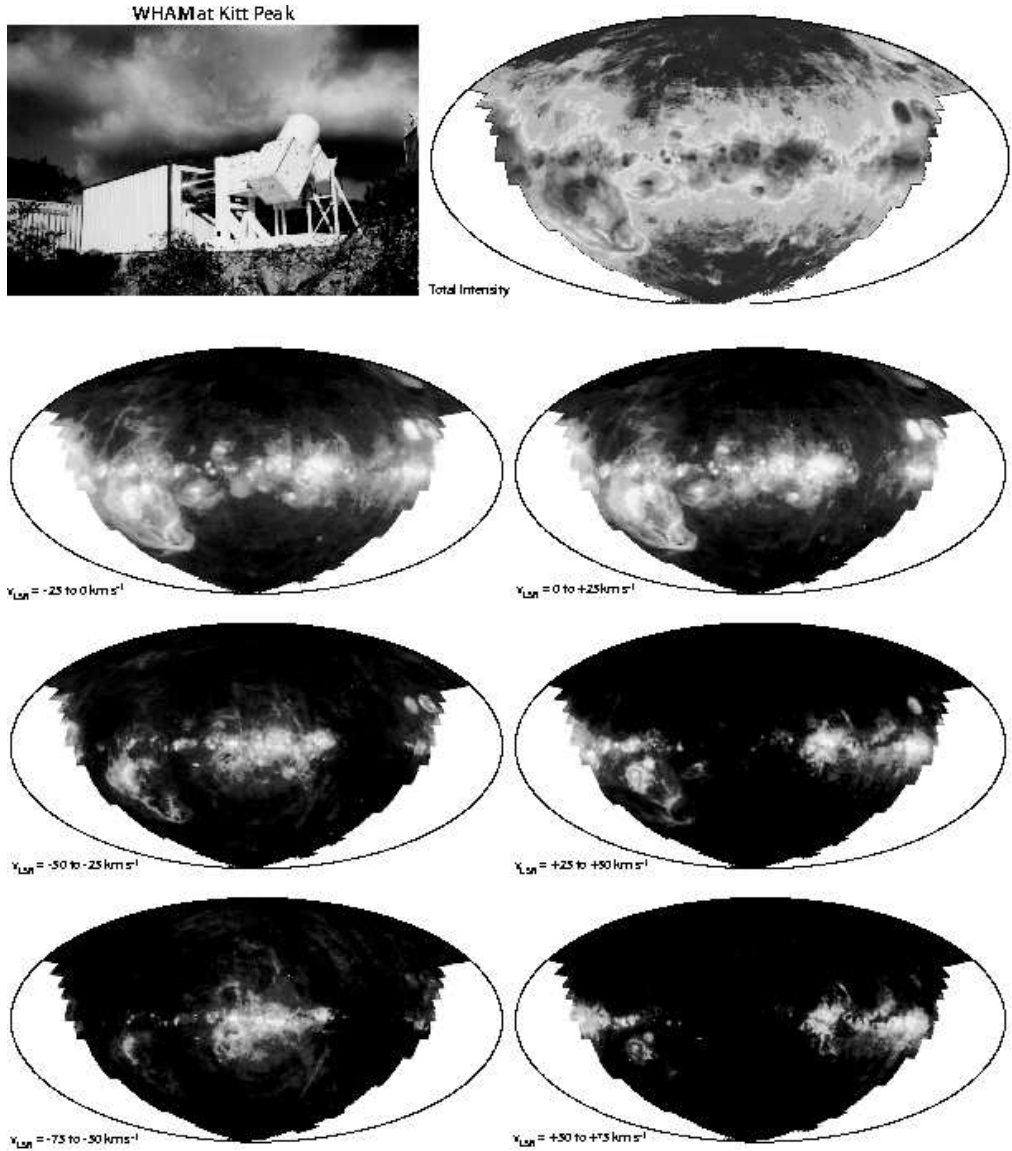}
    \caption{The WHAM facility at Kitt Peak plus H$\alpha$ total intensity
      and velocity interval maps from its recently completed Northern
      Sky Survey, revealing for the first time the distribution and
      kinematics of the diffuse H~II over the sky. All maps are
      centered at $\ell = 120\deg$. These data have been released to
      the community and are available at
      \texttt{http://www.astro.wisc.edu/wham/}.}
    \label{fig:survey}
\end{figure}

\section{Mapping the Excitation and Ionization State of the Gas}

With the H$\alpha$ survey providing a picture of the overall distribution
and kinematics of the warm ionized medium, the detection and study of
other emission lines can be used to probe the physical conditions within
the gas.  One of the outstanding questions is the source of the ionization
and heating.  Valuable clues are contained in the emission line spectrum,
which is characterized by high [N~II] $\lambda$6584/H$\alpha$ and [N~II]
$\lambda$6716/H$\alpha$ and low [O~III] $\lambda$5007/H$\alpha$ and He~I
$\lambda$5876/H$\alpha$ line intensity ratios relative to the ratios
observed in bright H~II regions around O stars (e.g., Rand 1997, 1998;
Haffner et al 1999). This implies ionization and excitation conditions in
the diffuse ionized gas that differ significantly from conditions in the
classical H~II regions. Not only are the conditions different, but they
vary significantly with distance from the midplane, from sightline to
sightline, and even from one velocity component to the next along a single
sightline (e.g., Haffner et al 1999; Collins and Rand 2001; see also Figs.
3 and 4 below).

To map these variations throughout the nearby spiral arms and to explore
how the observed differences in conditions are related to the structures
revealed by the WHAM-NSS and to the known sources of ionization, WHAM has
begun to map portions of the Galaxy in the [N~II] and [S~II] lines.  The
power of these observations is illustrated in Figure 3 below, from Haffner
et al (1999), who used WHAM to map [N~II] and [S~II] over a limited
($30\deg \times 40\deg$) region of the sky that sampled parts of the Local
(Orion) and Perseus arms.  These observations by WHAM confirmed for the
Milky Way, and have extended to much fainter emission, similar trends
noticed in emission line observations of other galaxies---namely, a
dramatic increase in [N~II]/H$\alpha$ and [S~II]/H$\alpha$ with increasing
distance $|$z$|$ above the midplane, with relatively small (but
statistically significant) variations in [S~II]/[N~II], which are
inconsistent with pure photoionization models (see Collins \& Rand 2001;
Otte et al 2001).

Haffner et al (1999) and Reynolds et al (1999) have pointed out that these
observations can be readily explained if the variations in the forbidden
line ratios are due primarily to variations in the electron temperature
($\Delta T_e \approx 2000$ K to 3000 K) of the gas rather than to
variations in the ionization parameter. This would naturally explain the
near constancy of [S~II]/[N~II], for example, since these optical
transitions of S$^+$ and N$^+$ have nearly identical excitation energies.
If true, variations in [N~II]/H$\alpha$ are tracing variations in the
temperature of the gas, and the small variations in [S~II]/[N~II] are
reflecting variations in the ionization state (i.e., S$^+$/S; see
discussion by Haffner et al 1999).  Recent follow up observations of the
Milky Way (WHAM observations using the temperature diagnostic [N~II]
$\lambda$5755/[N~II] $\lambda$6584; Reynolds et al 2001b) and other
galaxies (Collins \& Rand 2001; Otte et al 2001) have provided support for
this idea, although it may not apply in all cases (Martin \& Kern 2001).

\begin{figure}[htbp]
\plotone{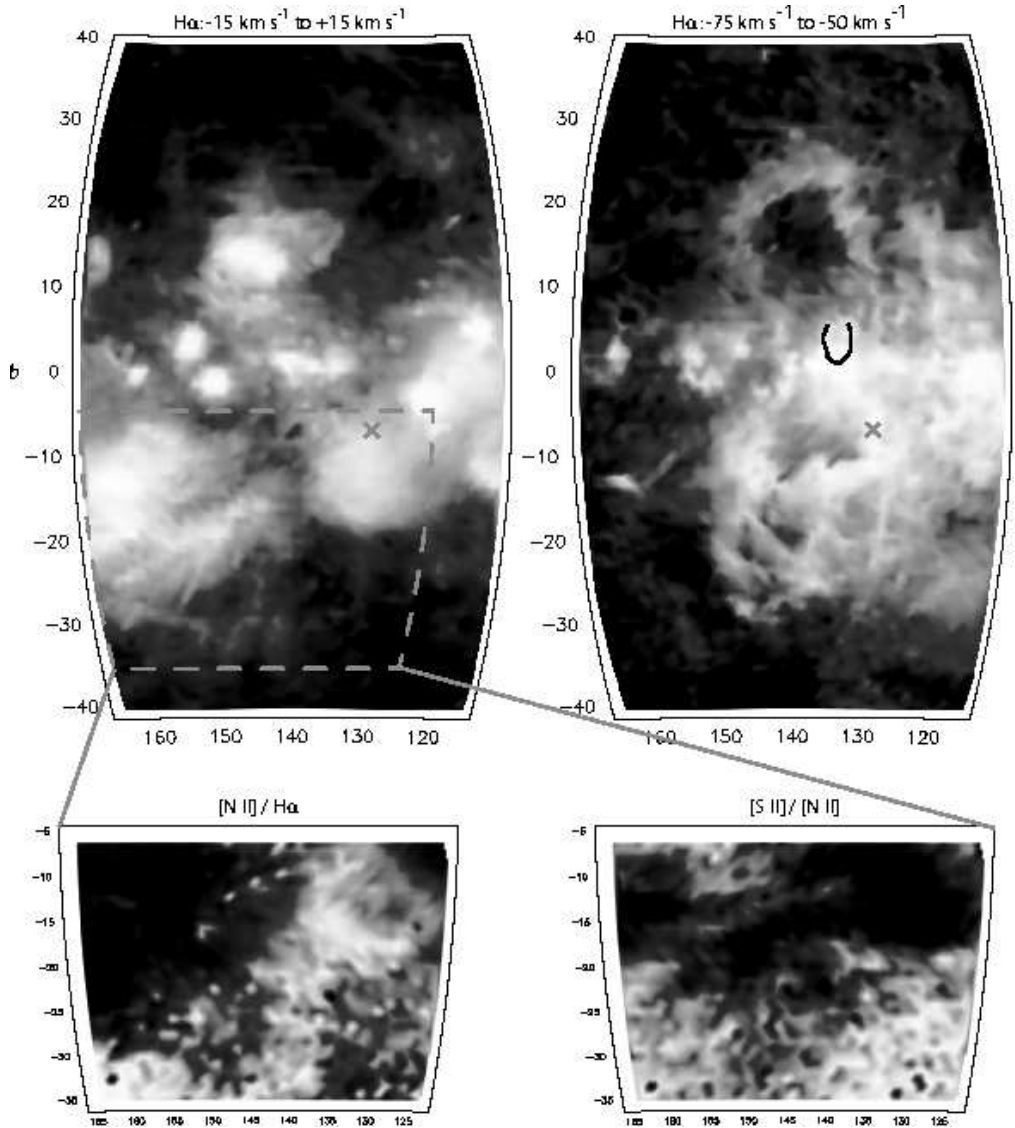}
    \caption{At the top of the figure are two WHAM velocity interval
      maps of a one-steradian portion of the sky showing emission from
      the Local Orion arm at V$_{lsr}$ = $-15$ to $+15$ km s$^{-1}$ and 
      from the
      more distant Perseus arm at V$_{lsr}$ = $-75$ to $-50$ km s$^{-1}$. 
      The large loop in the upper half of this second map is part of a
      superbubble that appears to have blown out into the Galactic  
      halo above the ``W4 chimney'' (the horseshoe schematic)
      associated with the Cas OB6 association (Normandeau et al 1996;
      Reynolds et al 2001a).  The ``$\times$'' denotes the
      sightline for the observations presented in
      Fig. 4.  Also shown for the region within the
      dashed boundary are $-15$ to $+15$ km s$^{-1}$ maps of 
      [N~II]/H$\alpha$ and
      [S~II]/[N~II], which are believed to trace variations in the
      excitation ($T_e$) and ionization (S$^+$/S) of the H$\alpha$ 
      emitting
      gas, respectively.} 
    \label{fig:ratiomaps}
\end{figure}

\begin{figure}[htbp]
\plotone{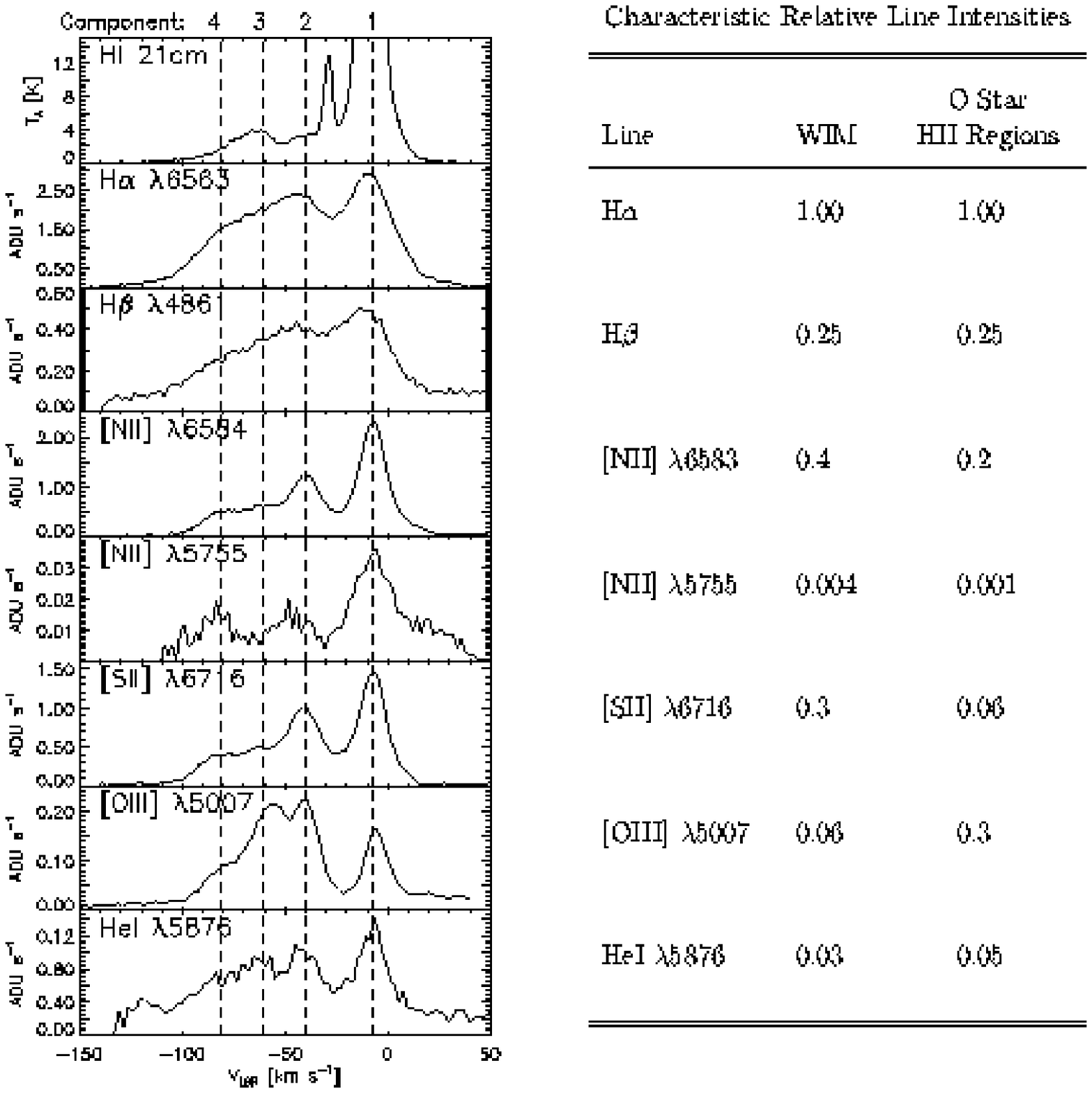}
    \caption{The H~I 21~cm spectrum from Leiden/Dwingeloo plus spectra of
      seven optical emission lines from WHAM toward $\ell = 130.0,\ b   
      = -7.5$ (denoted by an ``$\times$'' in
      Fig. 3). Except for the narrow H~I component at $-20$ km
      s$^{-1}$,
      each of the optically emitting clouds along this sightline
      appears to have a corresponding emission feature in the 21~cm
      spectrum.
      Note the variations in physical conditions along the sightline,   
      for example, the relatively high [O~III] in component 3 relative   
      to component 1 when compared to [S~II] or [N~II].}
    \label{fig:spectra}
\end{figure}
      
\section{Detecting Much Fainter Emission Lines}

WHAM also provides the opportunity to detect and study emission lines from
the diffuse interstellar medium that are too faint to have been detected
previously, such as [O~I] $\lambda$6300, He~I $\lambda$5876, [N~II]
$\lambda$5755, and [O~III] $\lambda$5007.  The intensities, widths, and
radial velocities of these lines contain unique additional information
about the ionizing radiation and conditions within the emitting gas, and
they can place strong constraints on theoretical models. For example,
these lines probe:
  
\begin{description}
      
\item[hydrogen ionization fraction:] The intensity of the [O~I]   
  $\lambda$6300 line relative to H$\alpha$, when combined with the 
  electron
  temperature (e.g., [N~II] $\lambda$ 5755/[N~II] $\lambda$6584; see  
  below), measures the hydrogen ionization fraction (i.e., the  
  ionization parameter) within the emitting gas. This diagnostic
  tightly constrains the possible mechanisms of ionization and   
  provides important information about the relationship between the
  H~II and the H~I within the diffuse interstellar medium
  (Domg\"orgen \& Mathis 1994). The first detections of diffuse,
  interstellar [O~I] (toward three sightlines with WHAM) appear to
  rule out the existence of warm, partially ionized H~I clouds, for
  example, and have revealed significant variations in [O~I]/H$\alpha$ 
  from
  cloud to cloud (Reynolds et al 1998a).

\item[spectrum of the ionizing radiation:] The He~I
  $\lambda$5876/H$\alpha$ recombination line ratio probes the hardness of 
  the ionizing spectrum.  The relatively weak He~I/H$\alpha$ intensity     
  ratios in the warm ionized medium compared to O star H~II regions
  (Tufte 1997; Heiles et al 1996; Rand 1997) imply either a
  reprocessing of the radiation from O stars or a significant ``new''
  source of soft Lyman continuum photons that has not been recognized
  (B stars?  See Cassinelli et al 1995).
      
\item[electron temperature:] The ratio [N~II] $\lambda$5755/[N~II]
  $\lambda$6584 is an unambiguous electron temperature diagnostic for
  ionized regions (Osterbrock 1989).  The $\lambda$5755 emission has  
  been detected by WHAM in multiple velocity components toward a
  diffuse background sightline (see Fig. 4),  
  revealing that the electron temperature in the diffuse H~II is
  approximately 2000 K warmer than that in bright, classical O star
  H~II regions (Reynolds et al 2001b). This result lends strong
  support for the existence of an additional heating source within the
  low density gas (\S 1.1 above).  

\item[state of higher ions:] Observed variations in the weak [O~III]
  $\lambda$5007 line relative to lines from ions of lower ionization 
  state are a sensitive probe of changes in the state of the rarer,   
  higher ions in the gas (e.g., Rand 1997).  For example, toward
  130$\deg, -7\fdg5$ (Fig. 4), [O~III]/[S~II]
  varies considerably from one velocity component to the next.
  Comparisons with the [N~II] $\lambda$5755 and [N~II] $\lambda$6584 
  profiles indicate that the relatively strong [O~III] in 
  velocity component 3
  and weak [O~III] in component 1 cannot be due to temperature
  differences, but rather must reflect significant variations in the
  abundance of the O$^{++}$ ion among the different clouds.
  
\end{description}
  
\subsection{The Relationship between H~I and H~II in the ISM, IVCs, 
and HVCs}

Along sightlines away from the Galactic midplane, there appears to be a
generally close relationship, both kinematically and spatially, between
the diffuse H~II and ``warm'' (broad component) H~I clouds, including the
distinct complexes of gas at intermediate and high velocities (e.g.,
Reynolds et al 1995; Haffner et al 2001).  In Figure 4, for example,
except for the narrow H~I component at $-20$ km s$^{-1}$, each of the
optically emitting clouds along the $\ell = 130.0,\ b = -7.5$ sightline
appears to have a corresponding emission feature in the 21~cm spectrum.  
This close relationship between the diffuse H~II and the warm H~I phase is
apparent in many other sightlines (e.g., Haffner et al 2001; Hausen et al
2002; Reynolds et al 1995).  Therefore, these emission line observations
impact not only our understanding of the ionized gas, but also provide a
new insight into the nature of H~I clouds. The intensity of the H$\alpha$
emission provide a measure of the ionizing flux incident on the cloud
(Tufte et al 1998), while the other fainter lines probe the spectrum of
the radiation and the properties of the cloud's associated H~II.

Although their existence has been known for many years, the origin of High
Velocity Clouds (HVCs) is still not clear (e.g., Wakker 2001). This is due
at least in part to the fact that, until relatively recently, HVCs could
be studied in emission only via the 21~cm line.  The detection of HVCs in
optical emission lines has provided a fresh new approach to these
enigmatic objects (e.g., Tufte et al 1998; S. Tufte, in preparation),
resulting in information about their distances, abundances, origin, and
the environment in which they are located (Wakker et al 1999;
Bland-Hawthorn \& Maloney 1999; Weiner \& Williams 1996).

\section{Exploring the Influence of Extinction on Interstellar Emission 
Lines}

Interstellar extinction is significant within about 5\deg--10\deg\ of the
Galactic equator, where the optical depth through the disk at H$\alpha$
reaches unity or greater.  To measure the influence of extinction on the
low latitude portion of the WHAM-NSS, we have begun mapping the Galactic
plane in H$\beta$.  Because the emission is kinematically resolved, these
observations make it possible to map the effect of extinction (i.e., the
H$\beta$/H$\alpha$ ratio) as a function of position along the line of
sight and to explore with numerical radiation transfer models the
absorption and scattering of the diffuse interstellar emission within the
Galactic disk (K. Wood, private communication; Wood \& Reynolds 1999).
Many low latitude WHAM-NSS spectra between longitudes 15\deg\ and 35\deg\
show emission out to radial velocities of +80 km s$^{-1}$ or greater,
indicating that WHAM can probe into the inner Galaxy to a distance of 4
kpc or more.
      
\section{High Angular Resolution (3\arcmin) Imaging}

A relatively recent optical--mechanical upgrade has given WHAM an imaging
capability, making it possible to obtain deep, high angular resolution
(3$\arcmin$), very narrow band images having a selectable band width from
20 to 200 km s$^{-1}$ (0.4 to 4 \AA) within its 1$\deg$ diameter beam. At
H$\alpha$, a signal-to-noise ratio of 9 is reached for 0.5 R per 3\arcmin\
pixel in a 25 minute exposure. Longer integrations times are required for
[N~II] and [S~II] images.  By revealing any structure within WHAM's 1\deg\
diameter beam, such imaging observations can be used to interpret properly
interstellar emission line and absorption line studies toward the same
sightline.  In addition, these high spatial resolution observations can
probe the small scale structure of filaments and ``WHAM point sources''
(large, low surface brightness planetary nebulae?) discovered in the
survey, as well as regions that overlap with high angular resolution 21~cm
maps (e.g., Arecibo and the DRAO CGPS).  ``WHAM deep fields'' in high
latitude directions may provide insight into the small scale structure (if
any) in the properties and kinematics of the diffuse ionized regions.

\section{Summary}

The development of high throughput, high spectral resolution Fabry-Perot
spectroscopy has established that faint, diffuse interstellar emission
lines at optical wavelengths contain a wealth of new information about the
interstellar medium that cannot be obtained through other techniques at
other wavelengths.  The WHAM H$\alpha$ survey plus the detection of
fainter nebular lines reveal a complexity not only in the structure and
kinematics of the warm ionized medium, but also in its excitation and
ionization conditions.  Studies of this weak emission have begun to shed
new light on the nature of the interstellar medium and the principal
sources of ionization and heating within the disk and halo of the Galaxy.

\acknowledgements

This work was supported by the National Science Foundation grant AST96
19424.  We are also grateful to S. L. Tufte, K. Jaehnig, N. R. Hausen, J. 
Percival, R. Pifer, B. Babler, M. Quigley, and T. Tilleman for their 
contributions to WHAM, the survey observations, and the survey data 
reduction.

\end{document}